\documentclass[11pt]{article}
\usepackage[utf8]{inputenc}
\usepackage{graphicx}
\usepackage{enumitem}
\usepackage{caption}
\usepackage{subcaption}
\usepackage{gensymb}
\usepackage{titling}
\usepackage[T1]{fontenc}
\usepackage{amssymb}
\usepackage{amsmath}
\usepackage{stackengine}
\usepackage{readarray}

\usepackage{bm}
\usepackage{cancel}
\usepackage{mleftright}
\usepackage[normalem]{ulem}
\usepackage[tiny]{titlesec}
\usepackage{bbm}
\usepackage[margin=1in]{geometry}
\usepackage{graphicx}
\usepackage{float}
\usepackage{caption}
\usepackage{subcaption}
\usepackage{cite}
\usepackage{natbib}
\usepackage{authblk}
\PassOptionsToPackage{hyphens}{url}\usepackage{hyperref}
\usepackage{setspace}
\title{Assessing the contagiousness of mass shootings with nonparametric Hawkes processes}
\author{Peter Boyd, James Molyneux}
\affil{Oregon State University, Department of Statistics}

\begin{document}

\maketitle

\begin{abstract}
    Gun violence and mass shootings are high-profile epidemiological issues facing the United States with questions regarding their contagiousness gaining prevalence in news media. Through the use of nonparametric Hawkes processes, we examine the evidence for the existence of contagiousness within a catalog of mass shootings and highlight the broader benefits of using such nonparametric point process models in modeling the occurrence of such events. 
\end{abstract}

\section{Introduction}

Gun violence in the United States is a national public heath crisis \citep{bauchner} with firearm homicide rates 19.5 times that of other high-income countries \citep{Grinshteyn}. Mass shootings in particular represent a phenomenon of interest in that these high-profile events with multiple, and occasionally numerous, victims generate large amounts of media coverage. Such media coverage may lead to both a contagion effect that may incite others to carry out similar acts as well as an imitation effect that may allow mass shooters to learn from those that preceded them \citep{Meindl}. Though the term \emph{mass shooting} lacks a specific, rigorous definition, the number of gun related incidences with multiple victims has become so common in the past two decades that research of these events has become a necessary component of public health studies in the United States \citep{dzau}. From 2000 to 2018, the US Federal Bureau of Investigation (FBI) recorded 277 active shooter incidents in which an individual shoots and kills (or attempts to kill) others in a public space, resulting in 2430 casualties \citep{fbi}. The FBI further notes that the number of such incidents is on the rise, with 69\% of these incidents occurring between 2010 and 2018. The need to address the contagion factor of these events, whereby a single mass shooting event inspires or is correlated with future mass shooting events, represents a fundamental question in the underlying mass shooting phenomenon.
\\

Previous research by \cite{jetter2018} proposed that the ideation and implementation of mass shootings are linked to media coverage of such events. 
A contagion factor was also previously found by \cite{towers2015} which used a self-excitation contagion model to quantify the degree to which previous events inspired future events. In their work, \cite{towers2015} model the increased probability of a mass shooting event occurring on day $t_j$ given a previous event occurred on day $t_i$, $t_i < t_j$, and the average duration of the contagion process $T_{excite}$ using an exponential probability distribution. That is, the probability of a new mass shooting event occurring sometime in the 24 hours of day $t_j$ is expressed as
$$ P(t_j|t_i,T_{excite}) = \int_{t_j-t_i-1}^{t_j-t_i} dx \frac{e^{-x/T_{excite}}}{T_{excite}}.$$
\cite{towers2015} then couple this probability model with a non-contagion related baseline number of events, $N_0(t)$, and a total number of expected secondary events, $N_{secondary}$, to compute an expected number of events, $N^{exp}$, on day $t_n$ expressed as
$$N^{exp}(t_n) = N_0(t_n) + N_{secondary} \sum_{i:t_i < t_n} P(T_n|t_i,T_{excite}).$$
\\

By leveraging methods from the self-exciting point process literature, we propose to improve upon these previous studies in a few distinct but important ways. First, by formulating the occurrence of mass shootings as a nonparametric Hawkes process we avoid having to make assumptions about how the contagion factor decays over time. Whereas the decay of the temporary excitation in space, time, or space-time is well-understood in some fields, such as in seismology where the occurrence of aftershocks has been shown to decay according to a power-law, the decay of a potential contagion factor in mass shootings is less understood. Thus, by expressing the contagion factor nonparametrically we avoid having to assume that the the decay of the contagion factor follows some prescribed probability distribution.
\\

Secondly, by using the EM-type model independent stochastic declustering (MISD) method of \cite{marsan2008}, we can estimate the probability that any individual mass shooting event was caused by a previous event or is a non-contagion related background event. This in turn allows for the estimation of the background rate of mass shootings which can then be expressed as the expected number of background events by taking the sum of the background event probabilities. Further, the modeling framework allows for the expected number of secondary events to vary according the heinousness of the crime as measured by the number of victims.
\\

This article is then organized as follows. In Section \ref{Methods}, both parametric and nonparametric self-exciting and Hawkes point processes will be introduced along with model-fit assessments. In Section \ref{Data}, we introduce the four mass shooting data sets which we utilize in the analyses seen in Section \ref{Results}. Significance of results, comparisons to other analyses, and suggestions for future work will be discussed in Section \ref{Discussion}.
\\

\section{Methods}\label{Methods}

A point process is a random collection of points $\{\tau_1, \tau_2,\ldots\}$ occurring in some metric space \citep{daley2004}. These points often occur in some temporal or spatio-temporal window where $t_i\in \mathbb{R}$ represents the temporal dimension of the $i^{th}$ point and $s_i\in \mathbb{R}^n$ represents the spatial coordinates of the $i^{th}$ point. In practice, $\mathbb{R}^n$ is often taken to be $\mathbb{R}^2$ or $\mathbb{R}^3$ so that the spatial coordinates land on some two-dimensional plane or three-dimensional space where the third dimension can be taken to be the depth of the point. For our purposes, we consider the occurrence of mass shooting events to be a collection of $n$ marked spatio-temporal points, $\{(t_i, x_i, y_i, m_i):i = 1, 2, \ldots, n\}$, such that $t_i \in [0, T]$ represents the time the event occurred with 0 and $T$ taken to be the start and end of the temporal window, respectively, and $(x_i,y_i) \in [-\infty, \infty]\times [-\infty, \infty]$ represents the spatial location of the event. The mark, $m_i$, of point $i$ is then some additional covariate information which we take to be the number of victims, excluding the perpetrator, of the $i^{th}$ mass shooting event. For the marks of the process, we  define the number of victims to be the number of individuals either killed or injured during the shooting. In defining the marks in this way, we intend to measure how events with different numbers of victims impacts the ability of an event to incite future events.
\\

In general, point processes are typically modeled via their conditional intensity function, $\lambda(t)$ or $\lambda(s,t)$ for time and space-time point processes, respectively. The conditional intensity is defined as the infinitesimal expected rate at which points occur given the history of the processes, $\mathcal{H}_t$. That is, we model the occurrence of points in time as
\begin{equation*}
\lambda(t|\mathcal{H}_t)  = \lim\limits_{\Delta t \to 0} \frac{E[N([t,t + \Delta t))|\mathcal{H}_t]}{\Delta t}
\end{equation*}
or in space-time as
\begin{equation*}
\lambda(s,t|\mathcal{H}_t)  = \lim\limits_{\Delta s, \Delta t \to 0} \frac{E[N\left((s,s+\Delta s) \times (t,t + \Delta t)\right)|\mathcal{H}_t]}{\Delta s \Delta t}
\end{equation*}
where $N(\cdot)$ is taken to be a counting measure \citep{daley2004}.
\\

In what follows, we introduce the self-exciting, or Hawkes, point process and then elaborate further on the estimation and evaluation of the nonparametric version of the processes.

\subsection{Hawkes and Self-Exciting Point Processes}

When the occurrence of a point causes the temporary elevation in the occurrence of future points nearby in time or space and time, we refer to such a process as a self-exciting point process. Foundational work in self-exciting point processes was done by  \cite{hawkes1971} who defined the conditional intensity as
$$
\lambda(t|\mathcal{H}_t) =\mu + \sum_{i:t_i < t} \nu(t - t_i).
$$
where $\mu$ specifies the background rate in which events stochastically occur in time and $\nu$ is the triggering function which governs the temporary self-excitation of events. A self-exciting point process can be categorized as a branching process, or a mathematical process in which a background event occurs and spawns additional offspring events, which can in turn have additional offspring of their own.
\\

The temporal Hawkes process was later extended to the spatio-temporal domain where the rate of events can be modeled not just at time $t$ but also location $s$. 
When considering spatio-temporal self-exciting point processes specifically, the conditional intensity function is defined as
$$
\begin{aligned}
\lambda(s,t|\mathcal{H}_t) &=\mu(s) + \sum_{i:t_i < t} \nu(s - s_i, t - t_i). \\
 \end{aligned}
$$
where $\mu(s)$ describes the background rate for the occurrence of points as a function of the spatial location and $\nu(s - s_i, t - t_i)$ again describes the excitation of the events.
\\

There are numerous applications for self-exciting point processes, most notably in seismology \citep{ogata1988, ogata1998}, social networks such as email chains \citep{email} or retweets on Twitter \citep{tweet}, criminology and gang related violence \citep{crime}, terrorism \citep{terrorism}, neuroscience \citep{neuro}, and the spread of epidemic diseases like Ebola \citep{disease}. In this paper, we focus specifically on the realization of self-exciting point processes as applied to mass shootings.

\subsubsection{Epidemic-type aftershock sequences}

Modeling the temporal occurrence of earthquakes with self-exciting point processes was first proposed by \cite{ogata1988} and extended to the spatio-temporal domain in \cite{ogata1998}. 
These epidemic-type aftershock sequence (ETAS) models are parametric models based on well-studied phenomenon of the temporal decay of aftershocks, via Omori-Utsu \citep{omori1894,utsu1961}, and the magnitude distributions of earthquakes, via Gutenberg-Richter \citep{gutenberg1944}.
\\
 
ETAS models consider the triggering function $\nu$ to be composed of three separable functions $g(t)$, $h(x,y)$, and $k(m)$ pertaining to time, space, and magnitude, respectively. That is, the conditional intensity function is defined as
$$
\lambda(x,y,t|\mathcal{H}_t) = \mu(x,y) + \sum_{i:t_i < t} g(t-t_i) h(x-x_i,y-y_i) k(m_i)
$$
The functions $g(t)$ and $h(x,y)$ then model how the conditional rate of events decays over time and space, respectively, while the function $k(m)$ describes the productivity of previous events based on their marks. In seismology, for example, where marks are taken to be the magnitude, or amount of energy released during an earthquake, events with a larger magnitude will be more productive at producing offspring than an event of smaller magnitude.

\subsection{Nonparametric Hawkes}

Whereas ETAS models are based on well-understood properties of seismic phenomenon to parameterize the components of the triggering functions, mass shootings are much less well studied and understood. 
For this reason, we use a nonparametric Hawkes model, first introduced in \cite{marsan2008} as the model independent stochastic declustering (MISD) algorithm, to study the occurrence and contagiousness of mass shootings.
This modeling framework allows for a causal structure to be calculated probabilistically using an iterative process to estimate the probability that an event was caused by a previous event, or conversely is a background event. These probabilities are then used to estimate the constant values of step-functions for each portion of the triggering function, also known as histogram estimators. 
For the histogram estimators, the differences in the pairwise times and locations of the events are computed and then each inter-event time and location is placed into a set of disjoint intervals or bins. 
Based on the probabilities of the iterative process, constant values are estimated for each interval in order to fit the model.
The histogram estimator for the marks works similarly except that the disjoint intervals are created based on the marks themselves and not the pairwise differences.
\\

Using a nonparametric Hawkes model to describe mass shootings then allows us to model the contagion factor of these events without making parametric assumptions about the shape and rate of decay. It also allows us to obtain estimates for the background rate of events directly based on the estimated probabilities of the model.
\\

Following the work of \cite{marsan2008}, we define 
\begin{equation*}
p_{ij} = \begin{cases} \text{probability event $i$ is triggered by event $j$,} & i > j \\ 
\text{probability event $i$ is a background event
    ,} & i = j \\
\text{0,} & i < j.
\end{cases}\end{equation*}
These probabilities can then be displayed as a lower-triangular probability matrix $P$ with 
\[
P = \begin{bmatrix} 
    p_{11} & 0 & 0 & \dots & 0 \\
    p_{21} & p_{22} & 0 & \dots & 0 \\
    p_{31} & p_{32} & p_{33} &  \dots & 0 \\
    \vdots & \vdots & \vdots & \ddots & 0\\
    p_{n1} & p_{n2} & p_{n3} & \dots & p_{nn} 
    \end{bmatrix}.
\]
Each row, $i$, of the probability matrix then describes the probability that event $i$ was caused by event $j$, $i > j$, or is itself a background event, $i = j$. Thus, each row of the probability matrix must sum to one.
\\

After initializing the probability matrix $P$ by setting each $p_{i,\cdot} = 1/i$, we iterate over the following sequence of steps until convergence has been achieved:
\begin{enumerate}
    \item Update the stationary background rate of the process by computing the expected number of background events based on the estimated probabilities from the $P$ matrix.
    \item Update the histogram estimators of the triggering functions for each disjoint space, time, or mark interval using the estimated probabilities from the $P$ matrix.
    \item Use the now updated background rate and triggering functions to update the probabilities that each event was either a background event or a child of a previous event.
\end{enumerate}
Convergence is achieved once the largest update to the entries of the probability matrix falls below some prescribed value $\varepsilon$. For a more detailed description of the algorithm, we refer readers to the article by \cite{Fox2015}.
\\

Standard errors for the histogram estimators of the triggering functions can also be computed to assess the variability of the estimates.
For the case of the temporal triggering function, $g(t)$, let $S_{\ell}$ denote a binomial random variable that represents the number of offspring in bin $\ell$, defined by parameters $\eta_t$, the true number of offspring, and $\theta_{\ell}^g$, the true probability a triggered event falls in bin $\ell$. We attain estimates of these parameters via 
$$\hat{\eta}_t = \sum_{i=1}^n \sum_{j=1}^{i-1} p_{ij} \quad \text{and} \quad \hat{\theta}_{\ell}^g = \sum_{B_{\ell}} p_{ij} / \hat{\eta}_t$$ 
for $p_{ij}$ equal to the triggering probability of the matrix $P$ upon convergence of the MISD algorithm and $B_{\ell}$ equal to the set of all events whose time differences fall within bin $\ell$ \citep{Fox2015}. As a result, we estimate the variance of the value $\hat{g}(t) = g_{\ell} = S_{\ell}/(\Delta t_{\ell}\eta_t)$ by 
\begin{equation*}
    \widehat{Var}(g_{\ell}) = \frac{(\hat{\theta}^g_{\ell})(1 - \hat{\theta}^g_{\ell})}{\hat{\eta}_t \Delta t^2_{\ell}}.
\end{equation*}
Standard errors for $\hat{k}(m) = k_{\ell}$ can be found similarly as
\begin{equation*}
    \widehat{Var}(k_{\ell}) = \frac{\hat{n}_t (\hat{\theta}^{k}_{\ell})(1 - \hat{\theta}^{k}_{\ell})}{(n_{\ell}^{mark})^2}
\end{equation*}
where 
$\hat{\theta}^{k}_{\ell} = \sum_{A_{\ell}} p_{ij}/\hat{\eta}_t$
for $A_{\ell}$ equal to the set of events whose marks fall within the $\ell^{th}$ marks bin and $N_{\ell}^{mark}$ equal to the number of events in bin $\ell$. 
\\

\subsection{Model Evaluation via Super-thinning}

Super-thinning \citep{clements2012} is a hybrid approach of two combined model evaluation techniques for point processes: residual thinning and superpositioning. 
For residual thinning, event $i$ is kept in the realized set of points, $\mathcal{S}$, with probability 
$b/\hat{\lambda}(s_i, t_i)$ 
for 
$b = \underset{(s, t) \in \mathcal{S}} \inf\left\{\hat{\lambda}(s,t)\right\}$ and removed from the data otherwise, where $\hat{\lambda}$ represents the estimated conditional intensity of a point in $\mathcal{S}$ \citep{Schoenberg2003}. 
Superpositioning meanwhile first simulates a point process with intensity 
$b - \hat{\lambda}(s, t)$, 
where 
$b = \underset{(s, t) \in \mathcal{S}} \sup\left\{\hat{\lambda}(s,t)\right\}$ and then superposes these points into the data \citep{bremaud}. 
For both thinning and superpositioning, the resulting residual process, $
\mathcal{R}$, will be a homogeneous Poisson process if and only if the model for the conditional intensity, $\lambda$, is correct. By using the hybrid approach of super-thinning, in that a point process is both thinned in areas of high conditional intensity and superposed points are included in areas of low intensity to form the residual process, the resulting set of points will have a higher power and lower volatility.
\\

The process then for super-thinning a point process, $\mathcal{S}$, is as follows:
\begin{enumerate}
    \item Thin $\mathcal{S}$ by retaining each point $(s_i, t_i)$ with probability $\text{min}\{b/\hat{\lambda}(s_i, t_i), 1\}$. 
    \item Simulate a point process with rate $\text{max}\{b - \hat{\lambda}(s,t),0\}$ at point $(s, t)$. 
    \item Combine the two resulting processes above to form the super-thinned residual process, $\mathcal{R}$. 
\end{enumerate}
The value of $b$ is used to adjust how much thinning or superposing takes place. 
Once a residual process is obtained, it can be examined for uniformity. If the model specification is correct then the residual process should have a uniform distribution throughout the time window.

\section{Mass Shooting Data}\label{Data}

Data availability on mass shootings is limited with no definitive collection of incidents reported by a public entity, in part due to the 1996 Dickey Amendment mandating that the injury prevention funds at Centers for Disease Control and Prevention (CDC) cannot be used to advocate or promote gun control [\citeyear{dickey}]. A 2018 spending bill clarified the language of the Dickey Amendment, allowing the CDC to research gun violence, which was believed to be barred by the amendment, while stipulating that government funds may not be used for gun control advocacy \citep{debonis}. Additionally, the United States government has no definition for a mass shooting but does define a mass killing as an incident in which a single perpetrator kills at least three people in a public space; this definition is consequently extended to the definition of a mass shooting by various entities that compile data for the purpose of studying mass shootings.
\\

\begin{figure}[h!]
  \centering
      \includegraphics[width=0.95\textwidth]{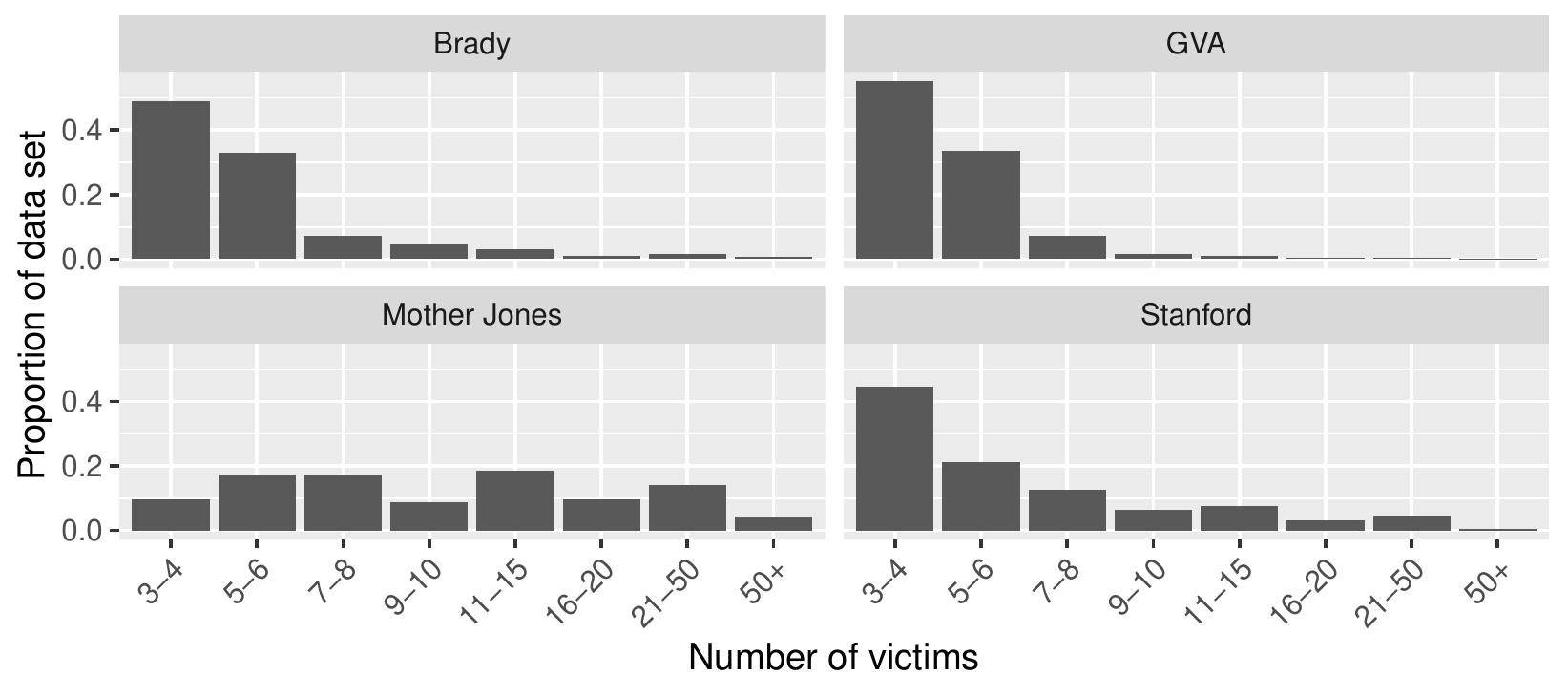}
  \caption{\label{fig:mag} Distribution of the number of victims for events in each data set.}
\end{figure}

Several private institutes and organizations have established publicly available data repositories that will be used in this study. Four data sets of mass shootings in the United States were utilized, and only events occurring in the continental United States were considered in analyses. The data sets differ in observation periods in addition to their definitions as to what constitutes a \emph{mass shooting}. Data compilation differences lead to large differences in total number of observations. Further discrepancies are acknowledged below and summarized in Table \ref{tab:data}. Plots containing the number of mass shootings per month are displayed in Figure \ref{fig:time}, with each plot displaying the same time window. Figure \ref{fig:mag} displays the distribution of the number of victims from events in each data set. 

\subsection{Brady: United Against Gun Violence}

The Brady Campaign (\url{https://www.bradyunited.org}) is a nonprofit group advocating for gun control and striving to end gun violence. The organization is named after James Brady, a cabinet member during the Ronald Reagan presidency who was shot during the assassination attempt on the president. Brady, left permanently disabled from the gunshot, became an advocate for gun control. The group has compiled data including incidents in which at least three people were shot or injured, but not necessarily killed. The data spans from February 2005 to January 2013, containing a total of 477 incidents. The Brady Campaign data set used in this article is also used in the \cite{towers2015} analysis to allow for comparison of results. Data can be accessed here: \url{https://journals.plos.org/plosone/article/file?type=supplementary&id=info:doi/10.1371/journal.pone.0117259.s002}.

\subsection{Stanford Mass Shootings in America}

The Stanford Mass Shootings in America data was compiled in an effort to create a comprehensive collection of mass shooting data in the United States. Incidents included involve three or more people shot, but not necessarily killed. The data ranges from August 1966 to June 2016 when maintenance and updates to the database were halted. We utilize data beginning in January 1999, with Columbine happening months later on April 20, 1999, to study the occurrence of mass shootings as a more modern phenomenon. The data originally contained 335 observations, but was reduced to 262 to reflect the altered starting date. Data can be accessed here: \url{https://library.stanford.edu/projects/mass-shootings-america}. 

\subsection{Gun Violence Archive}

Gun Violence Archive (GVA) (\url{https://www.gunviolencearchive.org}) is a nonprofit group that compiles records of gun related incidents in the United States. Incidents recorded involved four or more people shot but not necessarily killed. New records are updated in near real time, with data ranging from January 2012 until the present. While some data sets exclude events such as gang violence, GVA does not set any limiting terms to their definition of a mass shooting other than the number of individuals shot and killed, leading to a data set that contains a greater number of events. For events in which the perpetrator is killed or commits suicide during the shooting, GVA also differs from the other data sets in that the perpetrator is included in the number of total victims. 

\subsection{Mother Jones}

Mother Jones  is an investigative journalism organization that has compiled a collection of mass shootings under stricter criterion than others. With data ranging from 1982 until the present, Mother Jones initially recorded only incidents in which four or more people were killed. When the United States government redefined a mass killing to involve three or more people, Mother Jones followed suit, redefining the criterion for the database. For this analysis, the data will be reduced to events taking place on or after January 1, 1999. Data can be accessed here: \url{https://www.motherjones.com/politics/2012/12/mass-shootings-mother-jones-full-data/}.

\begin{table}[hbt!]
\begin{center} 
 \begin{tabular}{||c c c c c||} 
 \hline
 Dataset & Beginning Date & End Date & Definition & Observations \\ [0.5ex] 
 \hline\hline
Brady & February 2005 & January 2013 & 3+ killed & 477 \\ 
 \hline
Stanford & January 1999 & June 2016 & 3+ shot & 262\\
 \hline
Mother Jones & January 1999 & February 2020 & 3+ killed & 92 \\
 \hline
GVA & January 2012 & December 2019 & 4+ shot & 2024 \\ [1ex] 
 \hline
\end{tabular}
\caption{\label{tab:data}Summaries for each data set used in the analysis including the time window of data used in the analyses, definition of what constitutes a mass shooting, and the number of observations falling in the time window.}
\end{center}
\end{table}

\begin{figure}[h!]
  \centering
      \includegraphics[width=0.95\textwidth]{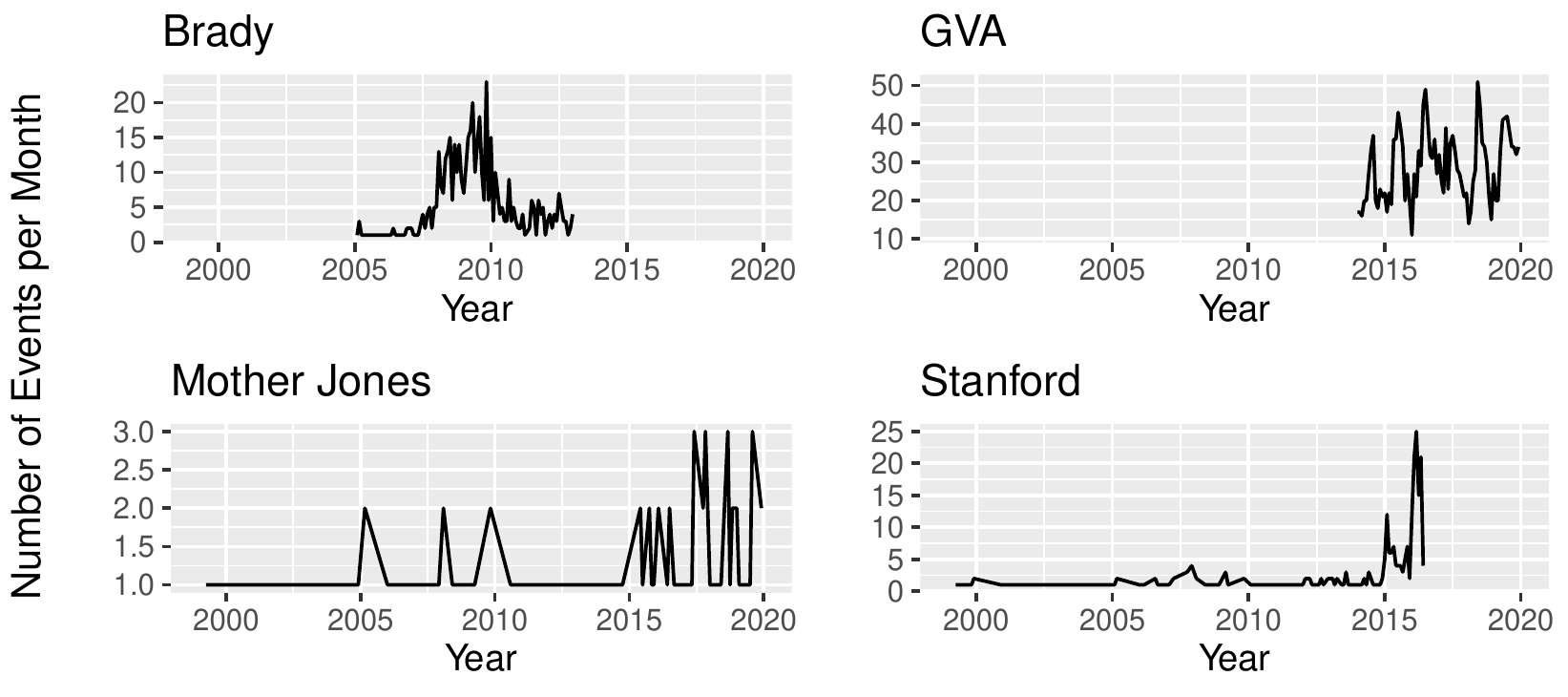}
  \caption{\label{fig:time} Monthly totals of the number of mass shootings for each data set.}
\end{figure}

\section{Results}\label{Results}

Nonparametric Hawkes processes were fit to each data set listed in Section \ref{Data} using the MISD algorithm to estimate their conditional intensity functions. Initially, the spatial triggering component was included in the conditional intensity function but was later dropped  as the spatial triggering component was found to have little to no effect in triggering subsequent events. The remaining results focus on the triggering of the temporal and mark components, $g(t)$ and $k(m)$ respectively. 
Intervals for the temporal triggering function were chosen to reflect natural breaks in the inter-event time differences, i.e. 2 weeks, 3 months, 6 months, +1 year, while the intervals for the marks triggering function were selected using quantiles to roughly allocate an equal number of events into each interval based on the number of victims. 
With discrete mark values, an exactly uniform division of events into quantiles could not be realized as certain values accounted for a large proportion of the data that would otherwise have spanned several quantiles, specifically in the GVA data in which roughly 55\% of incidents involved four victims.
\\

\begin{table}[hbt!]
\begin{center} 
 \begin{tabular}{||c c c c c||} 
 \hline
 Dataset  & Diagonal Mass & Background Rate & Number Offspring & Number 13 Day Offspring\\ [0.5ex] 
 \hline\hline
Brady & 10.39\% & 0.0168 & 0.8980 & 0.1913 \\ 
 \hline
Stanford & 28.42\% & 0.0117 & 0.7186 & 0.4140\\
 \hline
GVA & 34.87\% & 0.3484 & 0.6516 & 0.6146\\
 \hline
Mother Jones & 54.58\% & 0.0065 & 0.4592 & 0.0043 \\ [1ex] 
 \hline
\end{tabular}
\caption{\label{tab:results} Numeric summaries of implementing the MISD algorithm for each data set. Diagonal mass indicates the percent of the probability matrix $P$ that lies on the main diagonal. Background rate is the estimated background rate of the data catalog. Number of offspring is the estimated number of events that are triggered offspring of previous events, and number of 13 day offspring is the estimated number of offspring occurring within 13 days of an event. }
\end{center}
\end{table}

The diagonal mass of the probability matrix $P$, estimated background rate, average number of offspring events, and average number of offspring events occurring in the first two weeks are displayed in Table \ref{tab:results}. For most data sets, the majority of events are probabilistically treated as triggered events, with background events making up roughly 10\% to 55\% of observed mass shootings. The estimated background rate for the Brady, Stanford and Mother Jones data sets are estimated to be between 0.007 to 0.017 mass shooting events per day while the background rate for GVA is substantially larger with an estimated daily rate of mass shootings of 0.35. 
\\

For the Brady data set, the model estimated the expected number of offspring per mass shooting event to be roughly 0.90 events with 0.19 of those events, occurring in the first two weeks. This then implies that for an event in the Brady data, 21\% of the offspring events occur in the first two weeks with the remaining 79\% of events occurring sometime afterward.
The Stanford data had an estimated expected number of offspring per event of 0.72 with just over half, 0.41, of these events occurring in the first two weeks. 
The GVA data set had a slightly smaller overall expected number of offspring per event than Brady or Stanford with an expected number of 0.65 child events. However, the overwhelming majority, approximately 94\%, of the offspring events occurred in the first two weeks. 
Meanwhile, the Mother Jones data had the smallest expected number of offspring per events, 0.46 child events per mass shooting, yet 99\% of the child events occurred more than two weeks after the initial mass shooting.
\\

\begin{figure}[h!]
  \centering
      \includegraphics[width=0.95\textwidth]{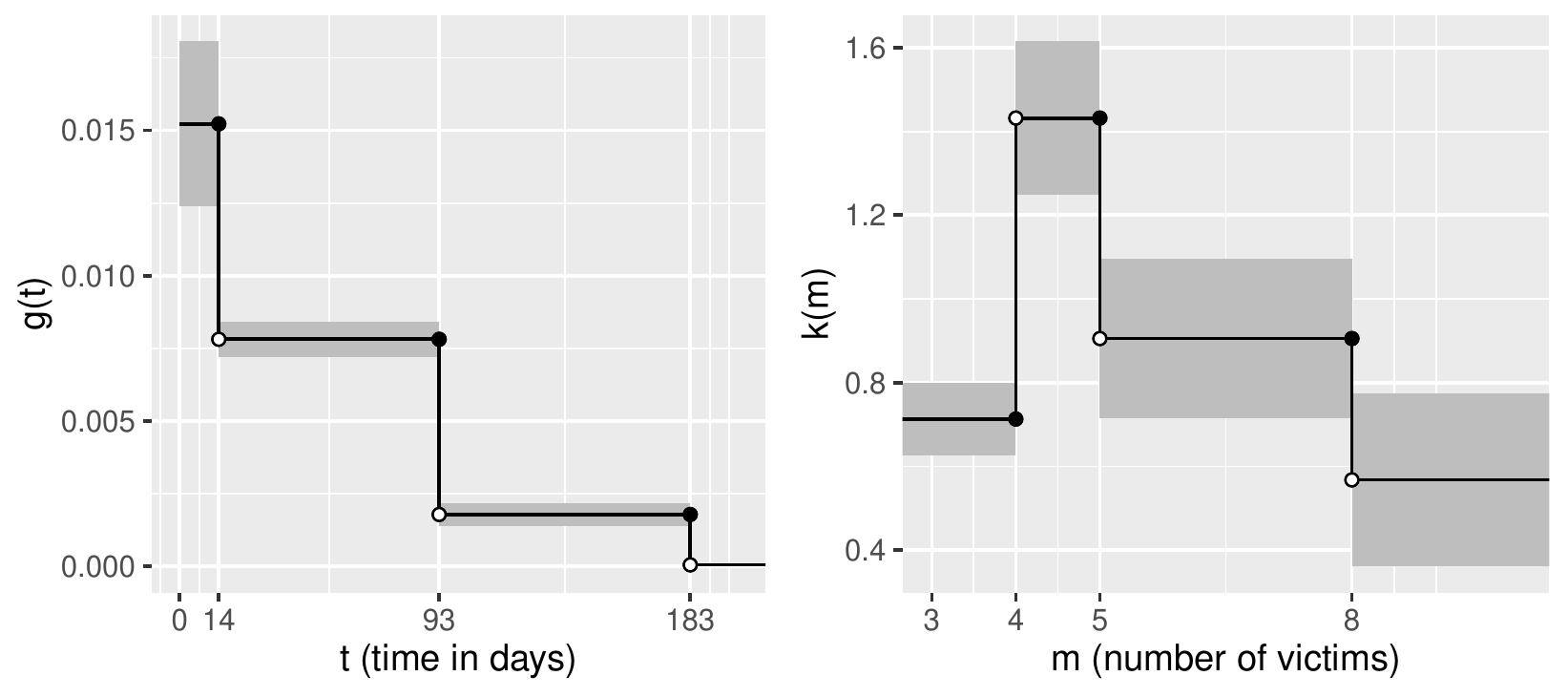}
  \caption{\label{fig:brady_trig} Brady Campaign triggering functions. In the figure on the left, values of the temporal triggering function are plotted over time, with the time bins used in the analysis shown on the $x$ axis. In the figure on the right, values of the marks triggering function are plotted over the marks (number of people injured). Standard error regions are shown in gray, and latter time bins with $g(t) \approx 0$ and the final mark bin is truncated in the figure. }
\end{figure}

\begin{figure}[h!]
  \centering
      \includegraphics[width=.95\textwidth]{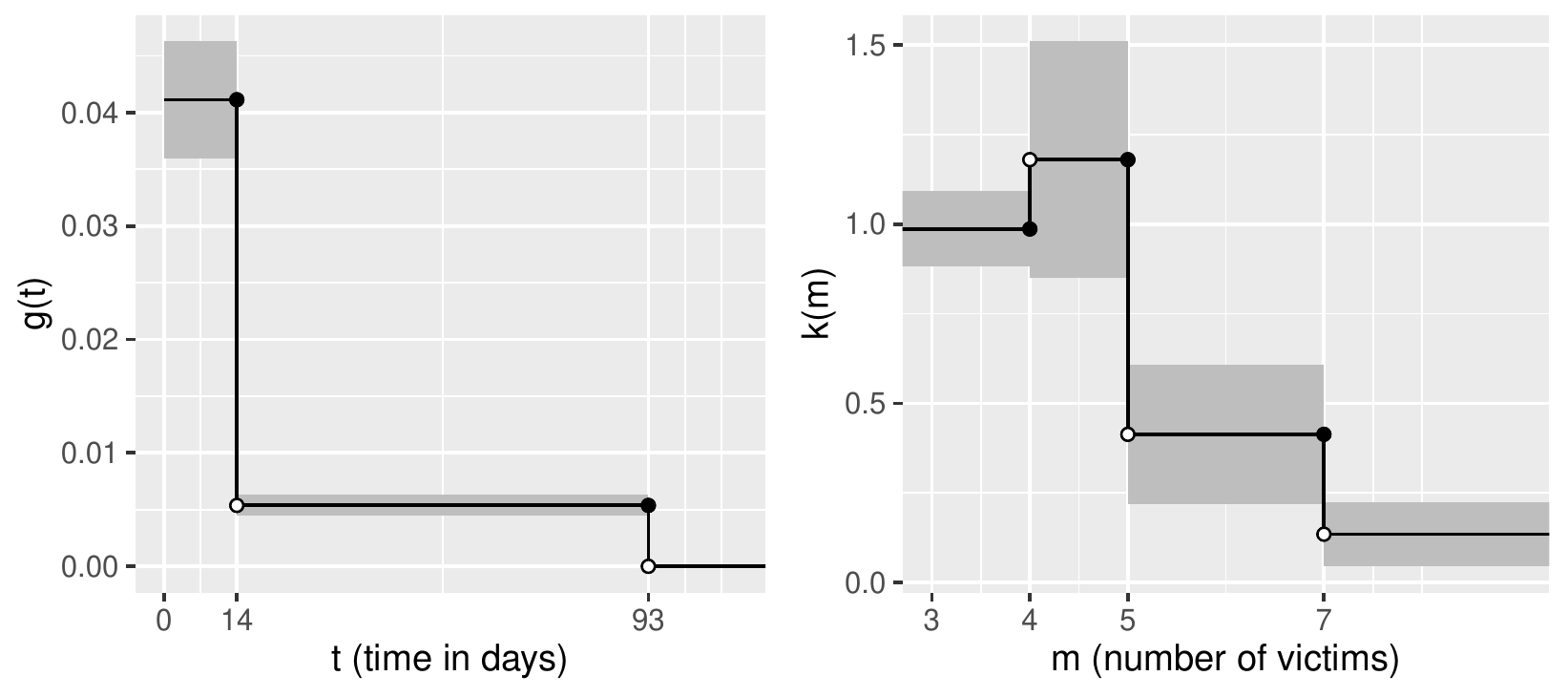}
  \caption{\label{fig:stanford_trig}Stanford triggering functions. In the figure on the left, values of the temporal triggering function are plotted over time, with the time bins used in the analysis shown on the x axis. In the figure on the right, values of the marks triggering function are plotted over the marks (number of people injured). Standard error regions are shown in gray, and latter time bins with $g(t) \approx 0$ and the final mark bin is truncated in the figure. }
\end{figure}

\begin{figure}[h!]

  \centering
      \includegraphics[width=0.95\textwidth]{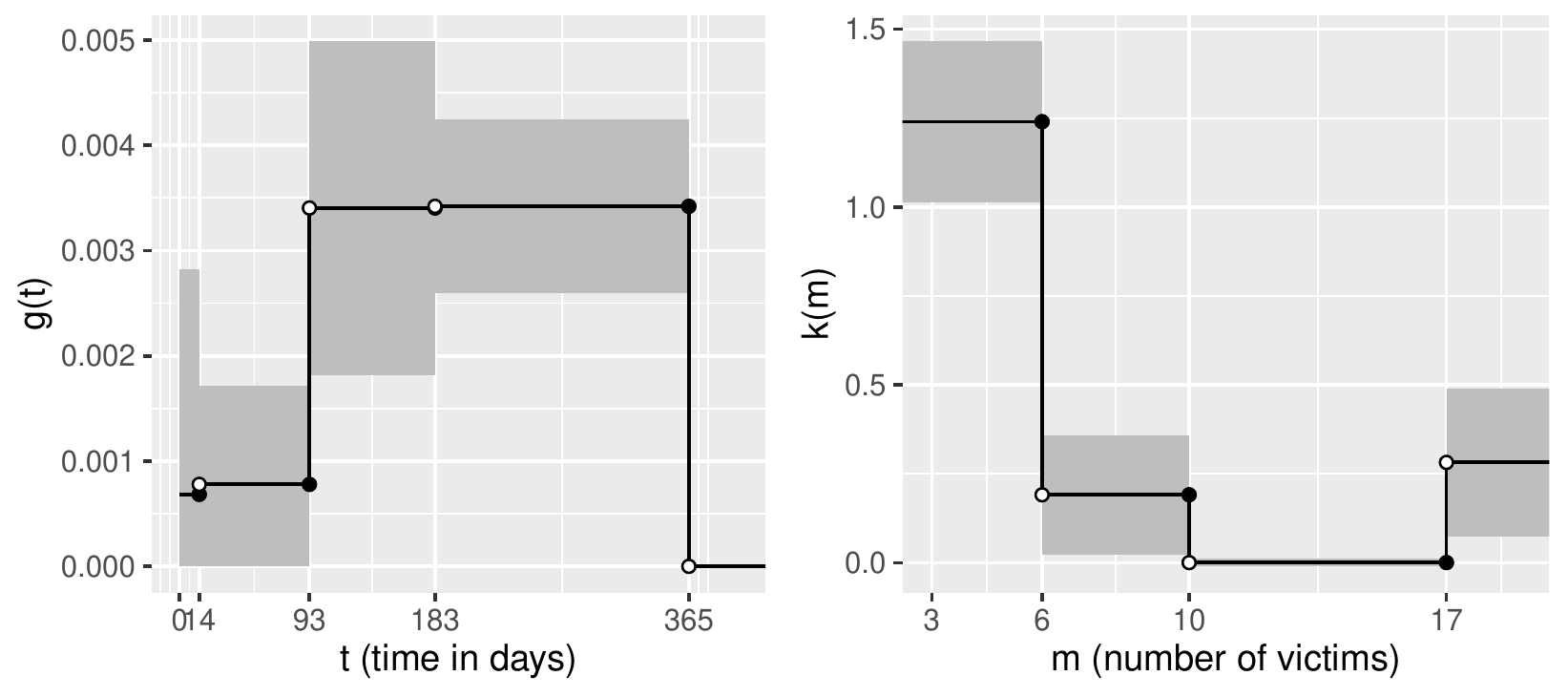}
  \caption{\label{fig:mj_trig}Mother Jones triggering functions. In the figure on the left, values of the temporal triggering function are plotted over time, with the time bins used in the analysis shown on the x axis. In the figure on the right, values of the marks triggering function are plotted over the marks (number of people injured). Standard error regions are shown in gray, and latter time bins with $g(t) \approx 0$ and the final mark bin is truncated in the figure.}
\end{figure}

\begin{figure}[h!]
  \centering
      \includegraphics[width=0.95\textwidth]{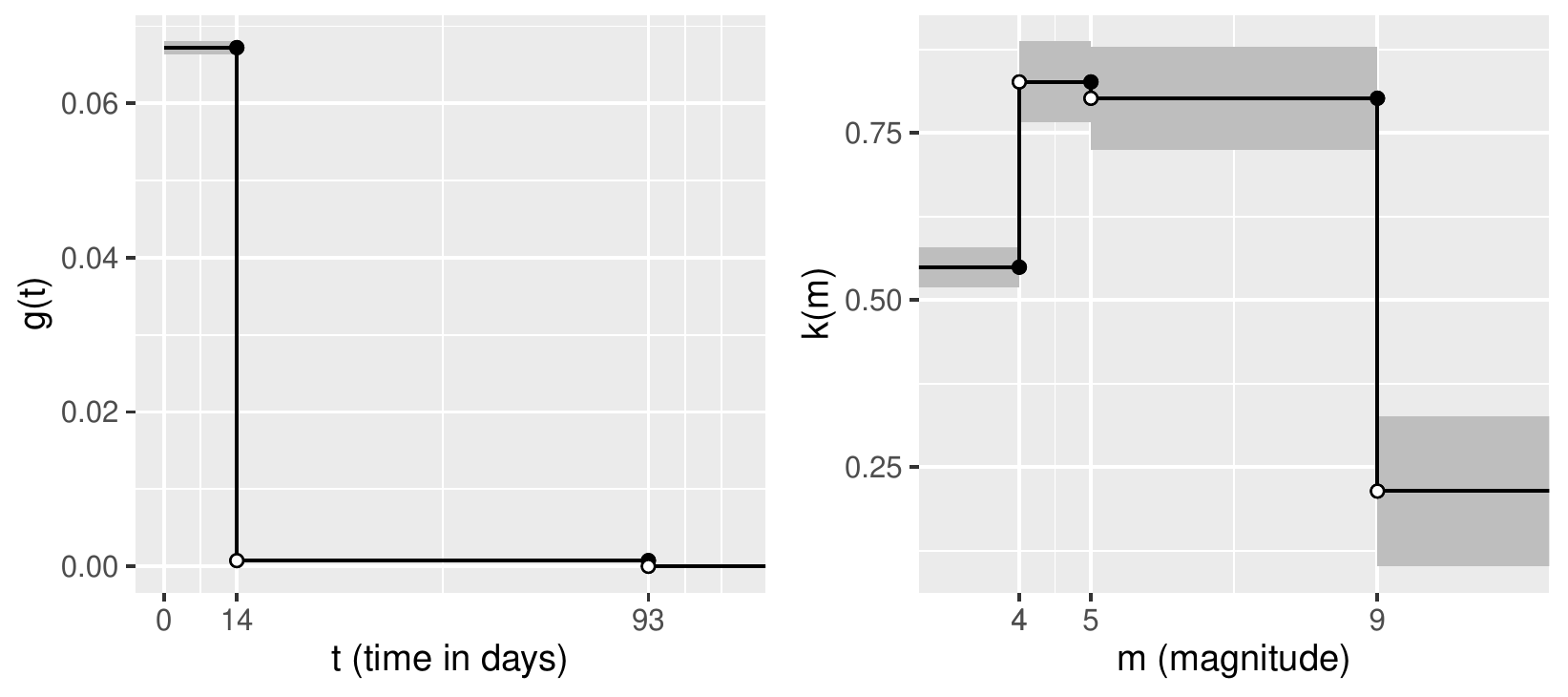}
  \caption{\label{fig:gva_trig}GVA triggering functions. In the figure on the left, values of the temporal triggering function are plotted over time, with the time bins used in the analysis shown on the x axis. In the figure on the right, values of the marks triggering function are plotted over the marks (number of people injured). Standard error regions are shown in gray, and latter time bins with $g(t) \approx 0$ and the final mark bin is truncated in the figure.}
\end{figure}

The estimated histogram estimators for the triggering functions of each data set are shown in Figures \ref{fig:brady_trig} - \ref{fig:gva_trig}. For each plot, the estimated constants of the histogram estimator step functions are shown as a horizontal line spanning the time or mark sub-interval for which the constant was estimated. 
The grey vertical bars then represent $\pm 2$ standard errors for each estimated constant of the histogram estimator. The standard error bars are truncated at zero to reflect only values that plausibly represent the phenomenon of interest. The temporal triggering functions, $g(t)$, are densities and thus the areas underneath the step function represent the probabilities of child event occurring over some time-span. The marks triggering functions, $k(m)$, represent productivity multipliers which increase or decrease the rate of triggered events based on the number of victims impacted in prior mass shootings. The $x$-axes of the temporal triggering functions are truncated as the functions tended towards zero as $t_j - t_i,$ for $j > i$, grew larger; $x$-axes of the marks triggering functions are truncated shortly after the final sub-interval as shown graphically.  
\\ 

In general, with the exception of Mother Jones, the value of the temporal triggering function, $g(t)$, monotonically decreases as $t$ increases to each subsequent time bin. For the Brady data, the decrease in the temporal triggering decreases more smoothly from roughly 0.0152 to 0.0078, to 0.0018 down to 0. For the Stanford and GVA data, the decay in the temporal triggering decreases much more drastically; from 0.41 down to 0.0054 down to zero for the first three time intervals in the Stanford data and from 0.067 down to essentially zero in the first two time intervals in the GVA data. For the Mother Jones data, the temporal estimates of the triggering are more volatile with estimates starting around 0.0007 and 0.008 for the first and second time interval, rises to around 0.0034 in the third and fourth intervals, then finally falls to zero. The Mother Jones data is also unique in that the estimated constants of the triggering function are much smaller in value than the other data sets.

For the estimated triggering functions of the marks for the Brady data, $k(m)$ had an estimated productivity of around 0.71 for the initial interval, and then increased to 1.43 for five victims, before falling to 0.91 for 6-8 victims and 0.57 for nine or more victims. The estimated mark triggering functions for Stanford and GVA contain the same pattern of an initial increase followed by two descending values. Stanford has an estimate of 0.99 for the initial bin and then jumped to 1.18 for five victims, before falling to 0.41 for 6-7 victims and 0.14 beyond 7 victims. GVA begins with at 0.55, increasing to 0.83 for 5 victims, then falls to 0.80 and 0.21 for 6-9 and 10+ victims, respectively. For the Mother Jones data, $k(m)$ also followed a less consistent form, with the highest value of 1.24 in the first bin before falling to 0.19 for 7 - 10 victims and 0.0007 for 11 - 17 victims before rising to 0.28 for 18 or more victims. The Stanford data yielded an estimated $k(m)$ that did not follow a monotone pattern, beginning at 0.99 in the first bin, increasing to 1.18 for five victims, then decreasing to 0.41 for six or seven victims, and 0.14 for larger numbers of victims.

\begin{figure}[h!]
  \centering
      \includegraphics[width=0.95\textwidth]{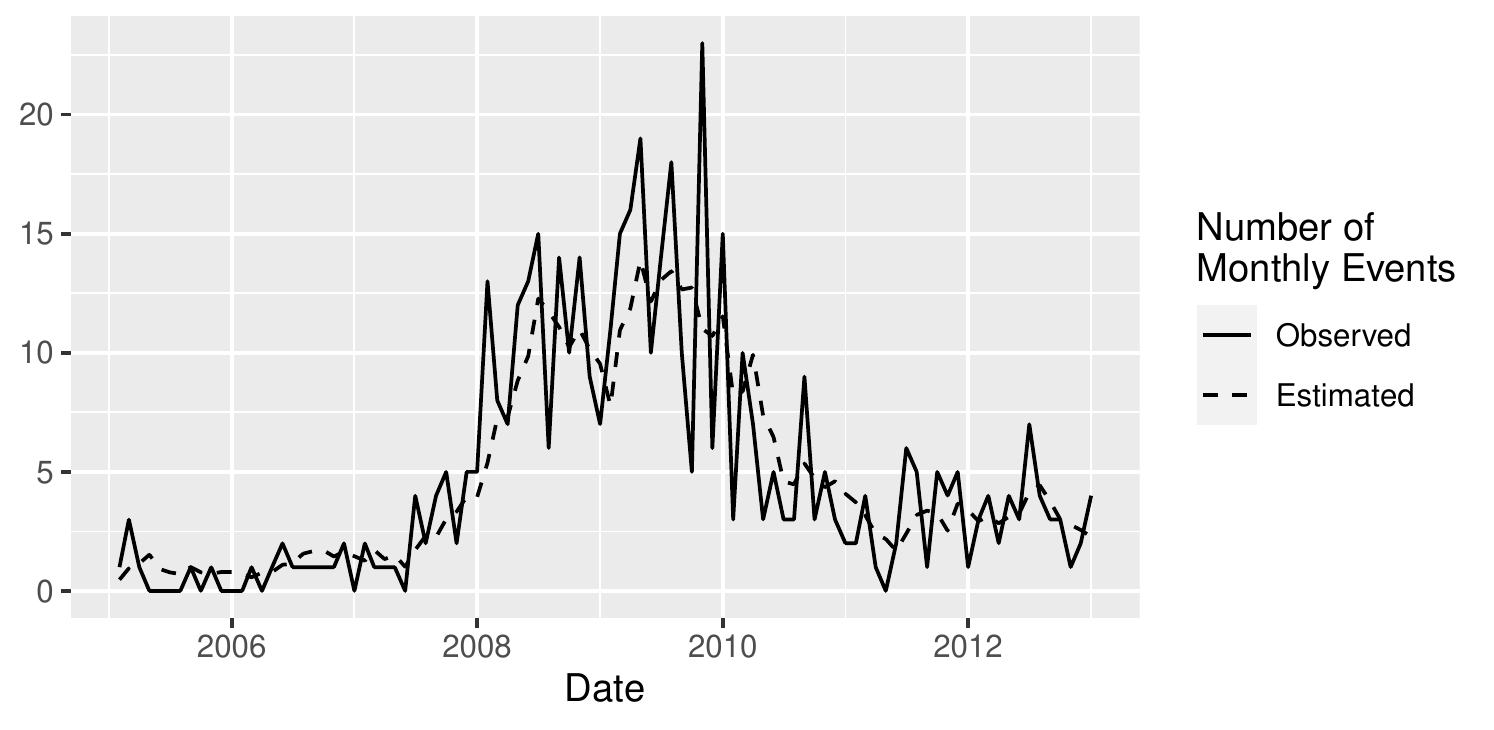}
  \caption{\label{fig:brady_ci}Brady Campaign conditional intensity plot. 
  The number of monthly mass shootings is plotted (solid line) over time. The median value of the estimated conditional intensity of the observed points is calculated for each month, multiplied by the number of days in each corresponding month, and plotted (dashed line) over time. 
  }
\end{figure}

\begin{figure}[h!]
  \centering
      \includegraphics[width=.95\textwidth]{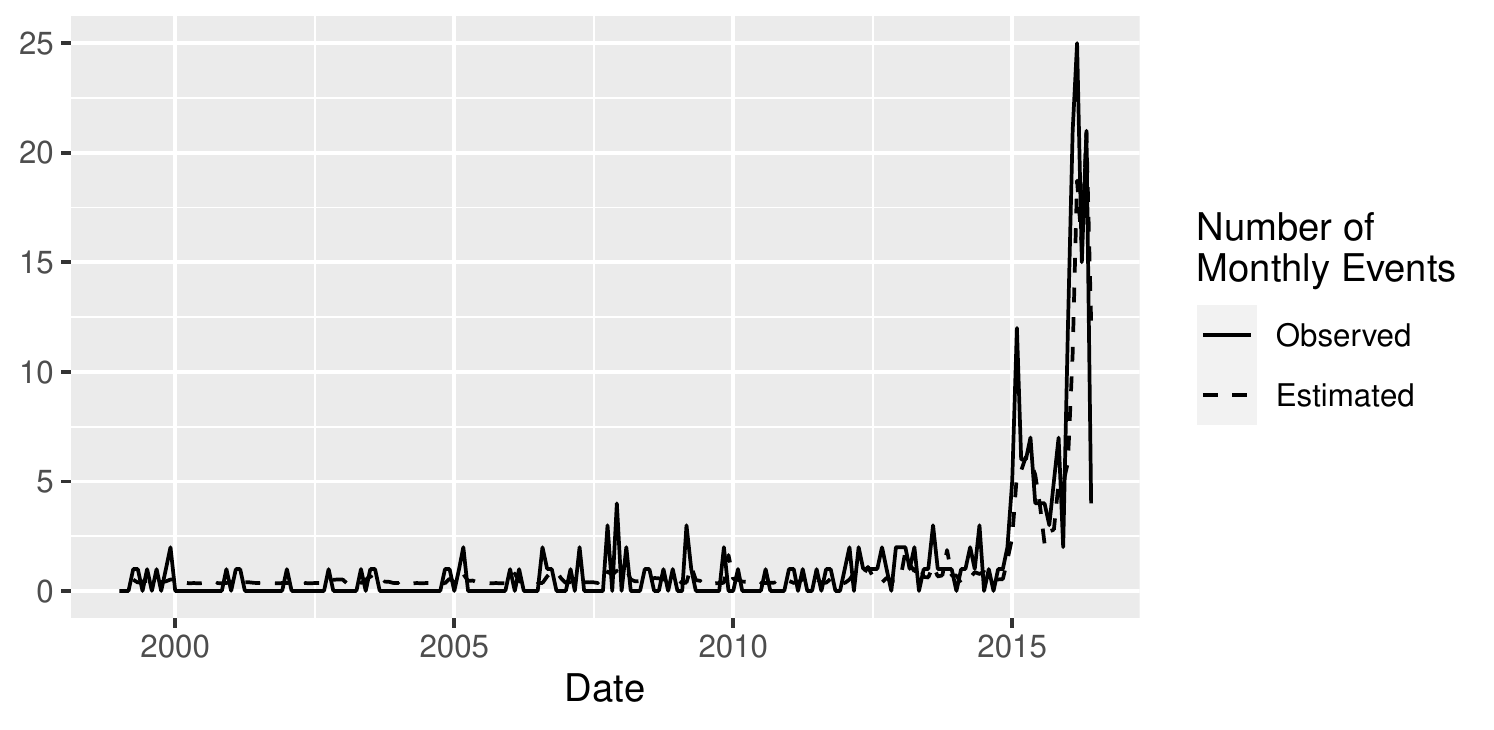}
  \caption{\label{fig:stanford_ci}Stanford conditional intensity plot. 
  The number of monthly mass shootings is plotted (solid line) over time. The median value of the estimated conditional intensity of the observed points is calculated for each month, multiplied by the number of days in each corresponding month, and plotted (dashed line) over time. }
\end{figure}

\begin{figure}[h!]
  \centering
      \includegraphics[width=0.95\textwidth]{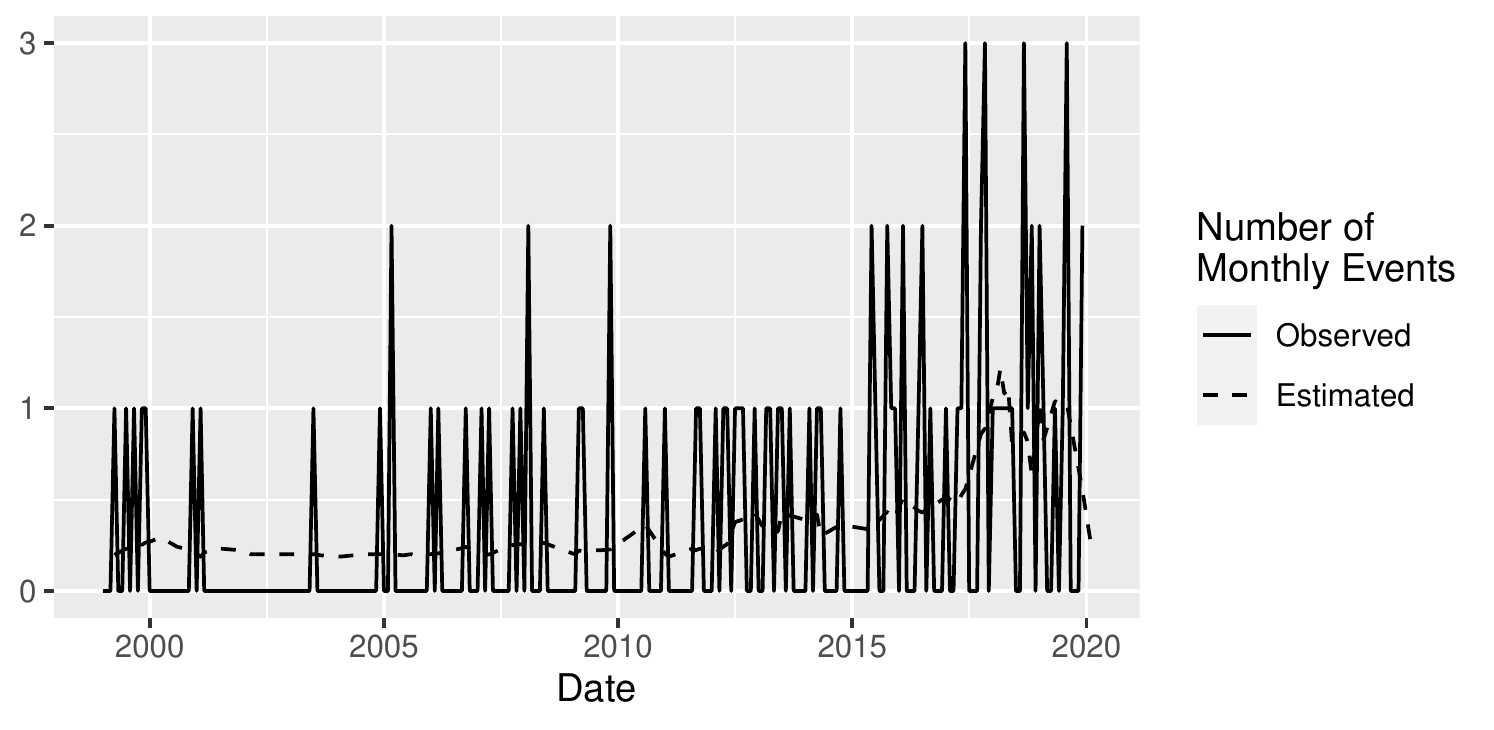}
  \caption{\label{fig:mj_ci}Mother Jones conditional intensity plot. 
  The number of monthly mass shootings is plotted (solid line) over time. The median value of the estimated conditional intensity of the observed points is calculated for each month, multiplied by the number of days in each corresponding month, and plotted (dashed line) over time.}
\end{figure}

\begin{figure}[h!]
  \centering
      \includegraphics[width=0.95\textwidth]{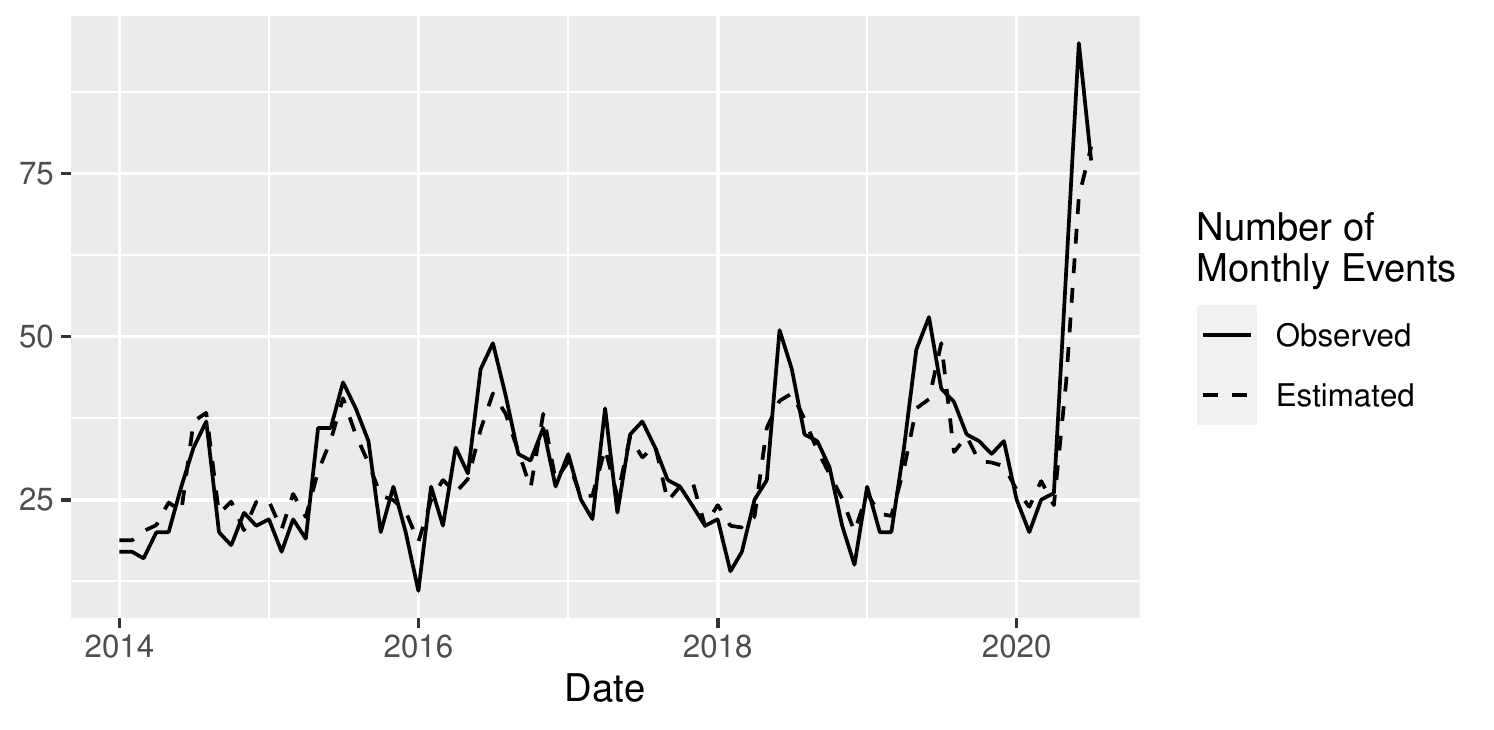}
  \caption{\label{fig:gva_ci} GVA conditional intensity plot. 
  The number of monthly mass shootings is plotted (solid line) over time. The median value of the estimated conditional intensity of the observed points is calculated for each month, multiplied by the number of days in each corresponding month, and plotted (dashed line) over time.}
\end{figure}

Figures \ref{fig:brady_ci} - \ref{fig:gva_ci} show the observed number of monthly mass shootings for each data source along with the estimated number of monthly shootings based on the models. The estimated values are computed by taking the median conditional intensity for each month and multiplying it by the length of the month.
The models appear to fit the data fairly well in that the estimated number of monthly mass shootings tends to follow the trends in the the observed values. 
The Mother Jones and Stanford data sets, Figures \ref{fig:stanford_ci} and \ref{fig:mj_ci} respectively, contain instances where no mass shooting events occurred over a sequence of consecutive months. For these months, the models tended to over-estimate the number of events as the model assumes a constant background rate.

\begin{figure}[h!]
  \centering
      \includegraphics[width=0.95\textwidth]{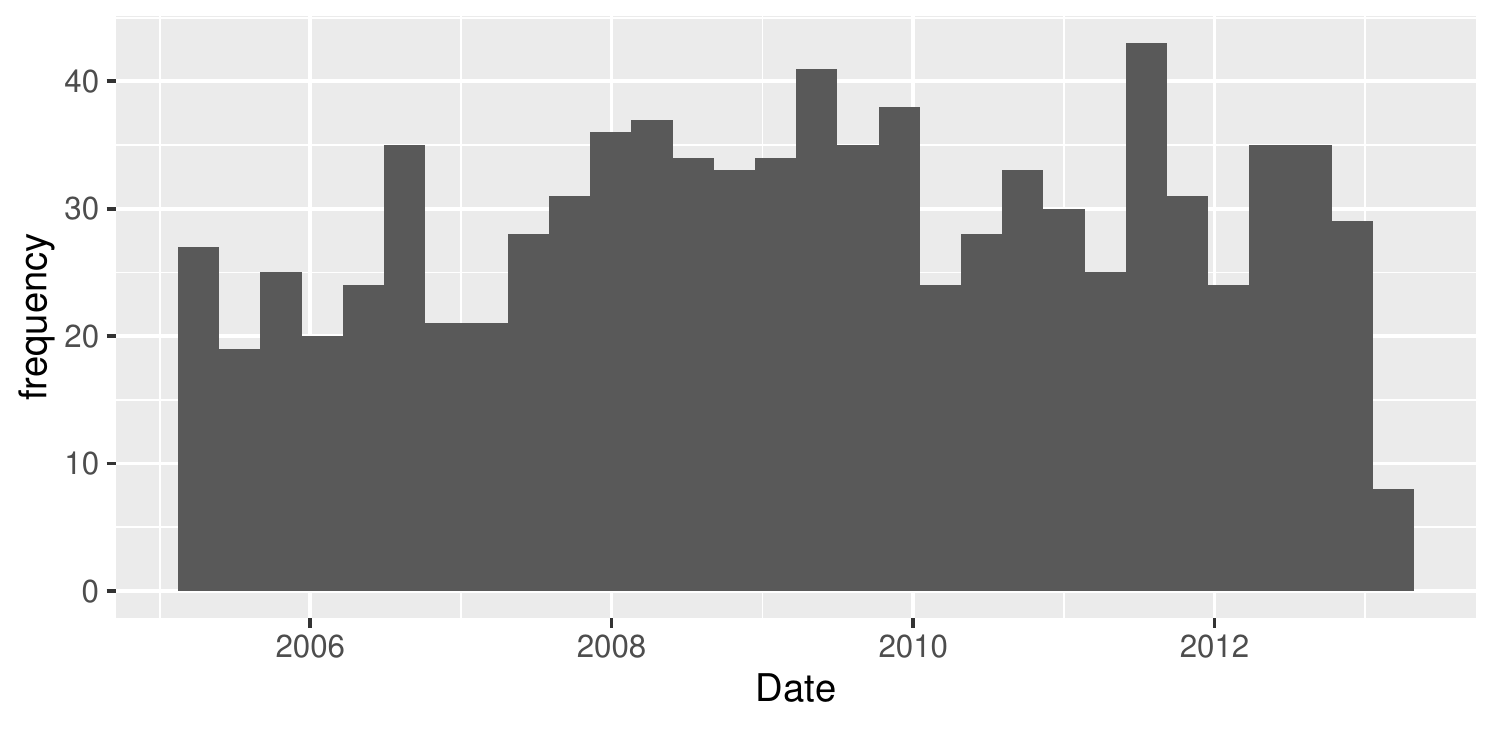}
  \caption{\label{fig:brady_hist}Brady Campaign histogram of super-thinned process. After super-thinning is implemented, the data are plotted over time, displaying the distribution of the super-thinned process. }
\end{figure}

\begin{figure}[h!]
  \centering
      \includegraphics[width=.95\textwidth]{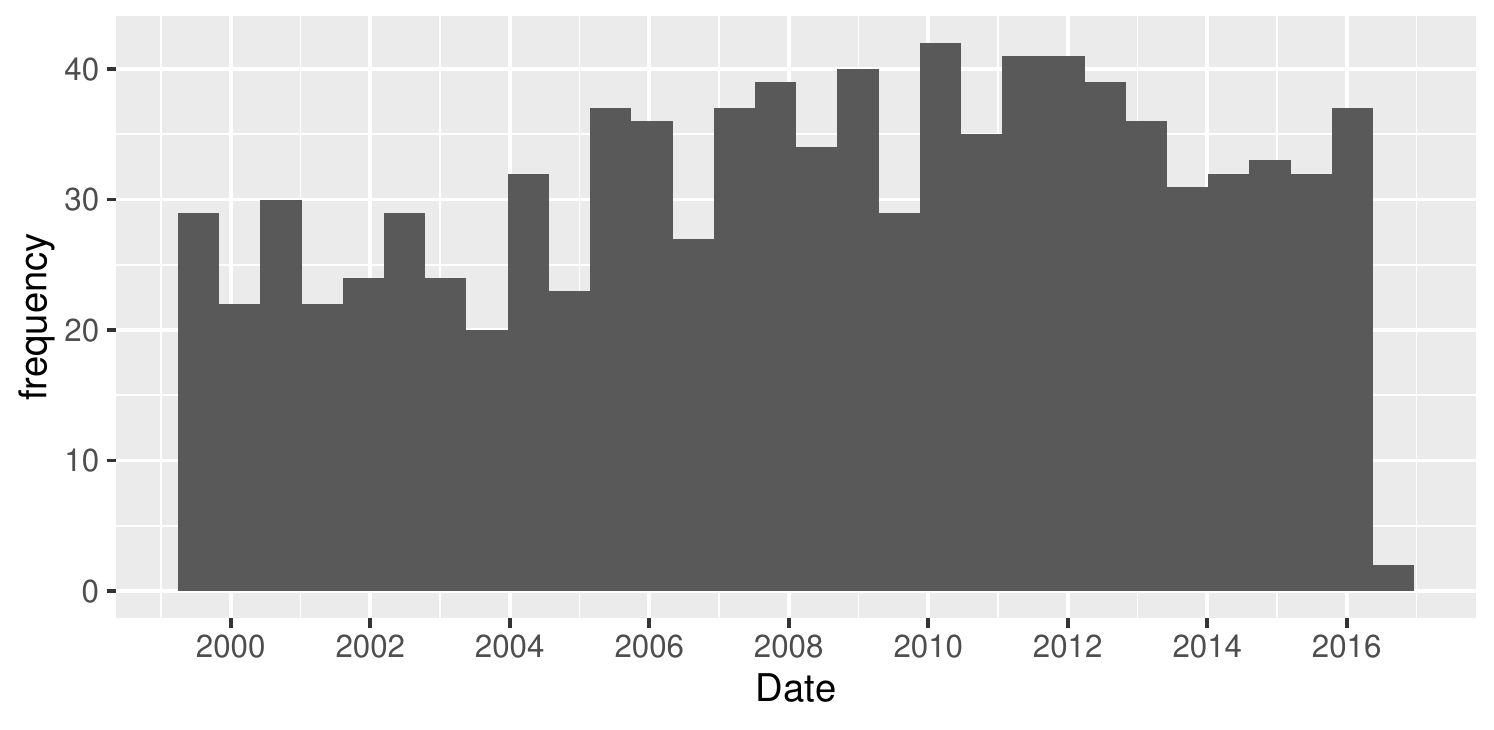}
  \caption{\label{fig:stanford_hist}Stanford histogram of super-thinned process. After super-thinning is implemented, the data are plotted over time, displaying the distribution of the super-thinned process. }
\end{figure}

\begin{figure}[h!]
  \centering
      \includegraphics[width=0.95\textwidth]{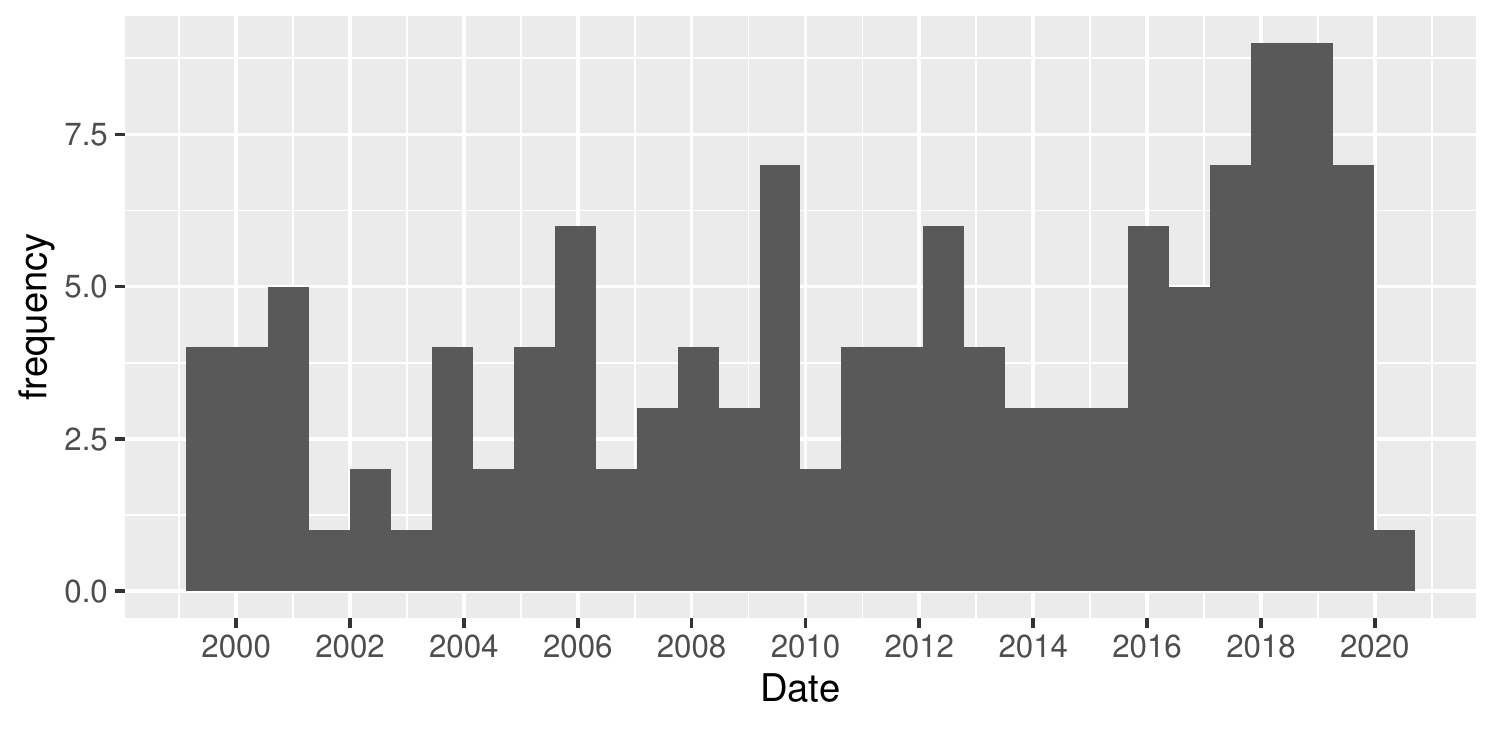}
  \caption{\label{fig:mj_hist}Mother Jones histogram of super-thinned process. After super-thinning is implemented, the data are plotted over time, displaying the distribution of the super-thinned process.}
\end{figure}

\begin{figure}[h!]
  \centering
      \includegraphics[width=0.95\textwidth]{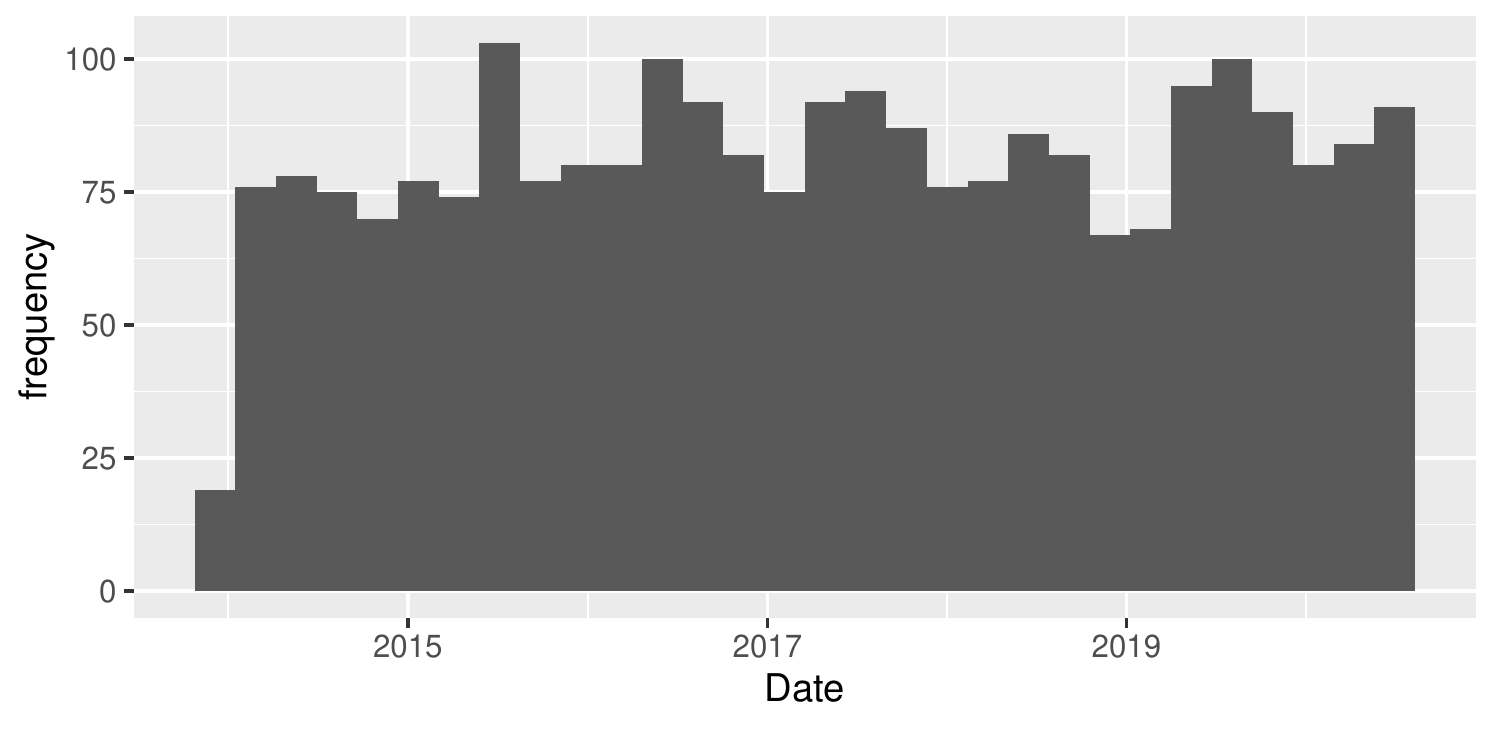}
  \caption{\label{fig:gva_hist} GVA histogram of super-thinned process. After super-thinning is implemented, the data are plotted over time, displaying the distribution of the super-thinned process. }
\end{figure}

Super-thinning was implemented to evaluate each model's fit to the individual data sets with tuning parameter, $b$, set to the median estimated conditional intensity for each source. To assess the overall fit of the model, the residual process for each data set is displayed as histograms in Figures \ref{fig:brady_hist} - \ref{fig:gva_hist}. 
If the model fits the data well, then we would expect the histograms to demonstrate a roughly uniform distribution throughout the entire time window. Of the four data sets, the estimated model for the Mother Jones data appears the least uniform in shape with substantial deviations throughout the time-window. The residual process for the GVA data appears the most uniform overall, though also with some deviations. The distributions of the Brady and Stanford deviation are somewhere in the middle with many time intervals appearing roughly uniform with some systematic deviations for certain time periods. The residual process for the Stanford data appears to have, in general, lower values prior to 2005 and slightly higher values in the years following, while the Brady residual process exhibits more of a unimodal distribution with a peak in values from 2008 - 2010.

\section{Discussion}\label{Discussion}

In this article, we investigate the contagiousness of mass shootings by treating the data as a marked self-exciting point process and analyze it through nonparametric Hawkes procedures. The contagiousness of mass shootings was previously studied by  \cite{towers2015}, reporting that each mass shooting will incite at least 0.30 new events brought on by an increase in probability of events that lasts for 13 days after an event. 
The self-excitation contagion model utilized in the Towers analysis requires several parametric assumptions including assuming a distribution for the decay of contagiousness, a constant number of secondary events, and the duration of contagion process. 
With little research on the contagiousness of mass shootings, circumventing the reliance on parametric assumptions through a nonparametric modeling framework is an important contribution to the study of this devastating phenomenon. 
\\

Through our nonparametric approach, we see evidence that events may produce higher numbers of offspring than previous results, with estimated number of offspring ranging from 0.59 to 0.86, as much as almost 3 times the value reported by \cite{towers2015} when using the same Brady Campaign data set.
We also note that a contagion effect exists after 13 days with expected number of offspring ranging from 0.03, based on the Mother Jones data, up to 0.60, with the GVA data set and 0.18 events in the Brady data.
The mean of these four values is 0.29, yielding an expected number of offspring within a 13 day period similar to the 0.30 reported by \cite{towers2015}. Also, similar to the results found in the Towers article, we noted no substantial spatial effect using the nonparametric framework.

In Figures \ref{fig:brady_trig} - \ref{fig:gva_trig},  the temporal histogram estimators tended to agree that the initial two-week period after a mass shooting event tended to have larger contagion effects compared to time periods after the initial two weeks, save for Mother Jones which had a temporal histogram estimator which was much more volatile. This volatility might not be entirely unexpected given that the Mother Jones data set had slightly more than one-third of the total number of observations compared to the next smallest data set but featured the longest time window of all the data sets. These factors then imply that very few of the pairwise time differences between events in the Mother Jones data fall in the shorter time intervals. The GVA data meanwhile is by far the largest data source with the shortest time-window and, as seen in Figure \ref{fig:gva_trig}, shows that nearly all of the contagion factor occurs in the first two-weeks. This is likely due to many of the pairwise inter-event time differences occurring relatively quickly after previous events.
\\

The triggering functions for the marks show much less consistency between the data sets but demonstrates the benefit of allowing the expected number of secondary events to vary depending on the size of the marks. The histogram estimator for the number of victims for the Brady campaign, Figure \ref{fig:brady_trig}, demonstrates that mass shootings with larger numbers of victims increases the productivity of those events in spurring future events. The Stanford data meanwhile, Figure \ref{fig:stanford_trig}, shows that events with between four to seven victims were more productive than larger events with greater than seven victims. A similar result was seen in the GVA data, Figure \ref{fig:gva_trig}. Again, the histogram estimator for the Mother Jones data, Figure \ref{fig:mj_trig}, is drastically different compared to the rest in that smaller events are more productive than larger events with high victim numbers. Furthermore, while the model appears to be finding some signal in regards to how the number of victims impacts the  productivity of mass shooting events to spur future events, it should be noted that there's a considerable amount of uncertainty in these estimates, as represented by the standard error bars, especially for the Stanford and GVA data. 
\\

\begin{figure}[h!] 
  \centering
      \includegraphics[width=0.95\textwidth]{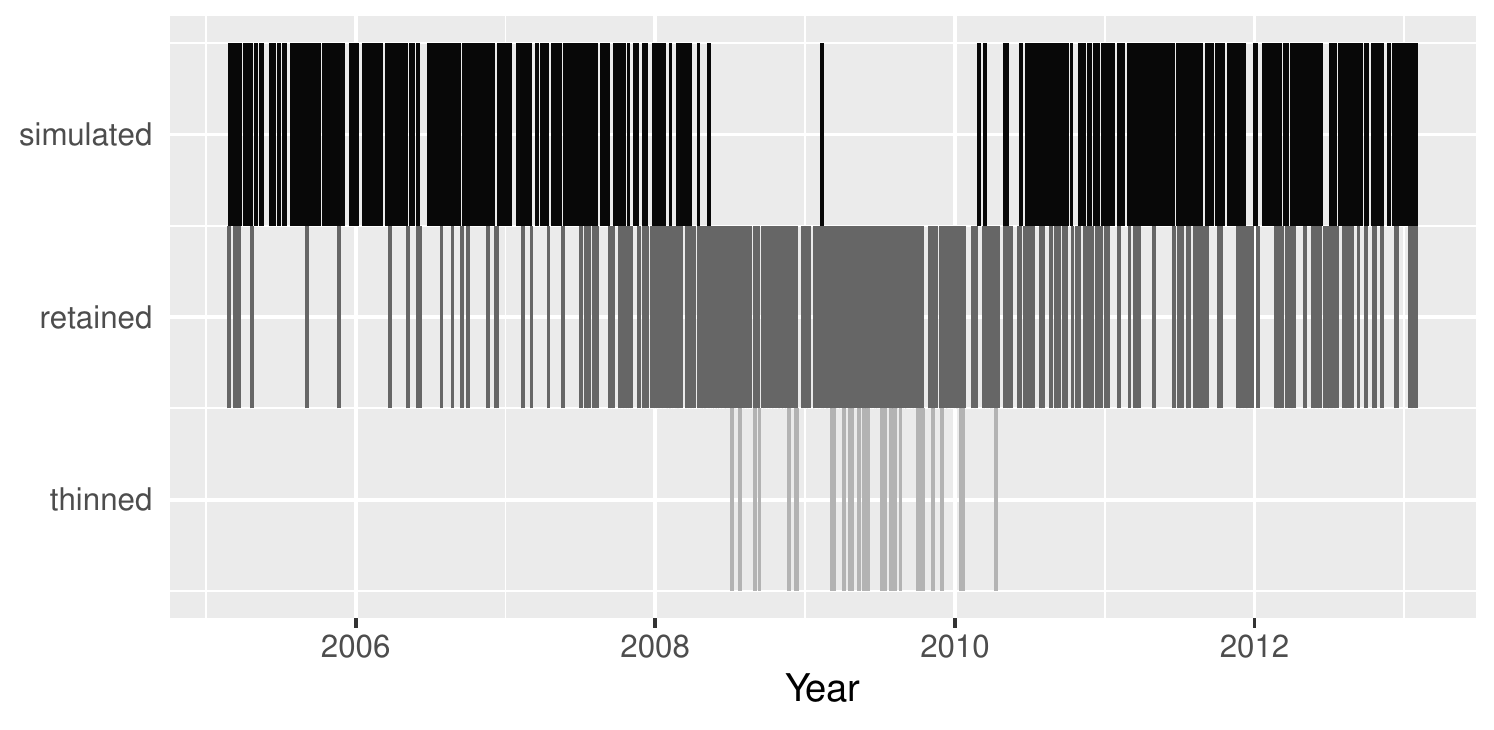}
  \caption{\label{fig:brady_st}Brady Campaign super-thinning plot. After super-thinning is implemented, events are plotted by their classification type over time, indicating events that were removed (thinned) from the data, events that were not removed (retained), and simulated events were superposed into the data (simulated). }
\end{figure}

Figures \ref{fig:brady_hist} - \ref{fig:gva_hist} show the results of super-thinning the point process models for the different data catalogs. By considering the uniformity of the super-thinned residual processes we can evaluate the overall fit of the models in that models that fit the data well should have a uniform appearance in the histograms. 
In Figure \ref{fig:brady_hist}, we observe that the residual process for the Brady model has a uni-modal appearance rather than the desired uniform distribution. 
Examining Figure \ref{fig:brady_st}, which shows the composition of the points for the super-thinned residual process for the Brady data, allows us to further investigate the unimodal distribution. The simulated lines at the top of the plot shows the points which were superposed while the retained lines show the points of the original process which were retained after thinning. The points which were thinned are then shown at the bottom of the plots.
From the figure, it is evident that the super-thinned process simulates events in areas of low intensity and removes events from areas of high intensity, but by simultaneously analyzing Figures \ref{fig:brady_hist} and \ref{fig:brady_st}, we see that a lack of sufficient thinning spurs departures from uniformity in the histogram. 
This lack of thinning then indicates that the model was not able to capture the full contagion effect present in the data.

In Figure \ref{fig:stanford_hist}, we observe an approximately uniform distribution, save a few spikes and falls, most notably at the end of 2010. Throughout the Stanford catalog, super-thinning appears to be performing as expected, despite the abrupt increase in the number of shootings that can be seen in Figure \ref{fig:stanford_ci}. Figure \ref{fig:gva_hist} also displays an approximately uniform distribution after super-thinning the GVA catalog.
\\

Figure \ref{fig:mj_hist} shows a non-uniform distribution of super-thinned residuals for the Mother Jones data. With so few events recorded in the Mother Jones data set, well-fitting models are more challenging to realize without adding further complexity to the model. In Figure \ref{fig:mj_ci}, the frequency of observed events appears to vary considerably over time, with 40\% of events occurring in only the last five years of the catalog. With such disparity in the frequency of events, fitting a single background rate for the entire process may oversimplify trends in the data; employing a nonconstant background rate may allow for a stronger representation of the data. 
\\

Varying data sets and definitions of mass shootings lead to seemingly inconsistent trends and results across analyses; more conclusive findings may be obtained with a more consistent definition of such events and better data collection methodologies. Comparisons of results across data sets can be difficult with data sources providing wildly different estimates; the Gun Violence Archive reports 2024 mass shootings over eight years, while the original Mother Jones data reports 118 incidents over nearly thirty-eight years. Although Brady and Mother Jones both define mass shootings as events in which three or more individuals are killed, the number of events in each data set are starkly different. The Stanford data set offers the well-fitting model but excludes data post 2016. The GVA and Mother Jones data, as shown in Figure \ref{fig:time}, have an upward trend in the number of mass shootings in later years; this trend may have also been evident in the Stanford data set had data collection been continued, potentially offering a broader understanding of mass shooting contagion, especially in later years. 
\\

Despite wildly different data and definitions, results are consistent in that a large percentage of mass shootings are probabilistically treated to be triggered events through the application of the MISD algorithm. Such findings support previously studied assertions that mass shootings may be motivated by a contagion effect spread through media.

\section{Conclusion}

In this article, we assess the the contagiousness of mass shootings using a nonparametric Hawkes process framework for a variety of data sources. This framework relies on fewer parametric assumptions than previous studies and detects a contagion effect which varies over both time and the number of victims. 
We also find that the level of contagion is contingent upon the data source used as no definitive catalog of data for mass shootings yet exists.
\\

Although the estimated conditional intensity for each process appears to closely mirror the true data process, more complex models with additional features may yield better fitting models in the future. Specifically, adapting a nonconstant background rate over time and/or a productivity function which is allowed to vary over time would allow future models to capture temporal changes to these two components. 
More complex models might also allow for the incorporation of meaningful spatial attributes or additional relevant covariates.
The models featured in this article then represent a baseline approach for the modeling of mass shootings as the nonparametric framework we implemented is extensible and able to benefit from innovations made in other fields and applications.

\bibliographystyle{plain}
\bibliography{ref}

\end{document}